\documentclass[iop]{emulateapj}

\usepackage{courier,graphicx,epsfig,color,colordvi,natbib,epstopdf}
\usepackage[normalem]{ulem}
\usepackage[colorlinks=true,urlcolor=blue,citecolor=blue,linkcolor=blue]{hyperref}
\usepackage[flushleft]{threeparttable}

\graphicspath{{./figure/}}

\begin{document}

\title{Penumbral brightening events observed in AR NOAA 12546}

\author{Mariarita Murabito\altaffilmark{1}, Salvo L. Guglielmino\altaffilmark{2}, Ilaria Ermolli\altaffilmark{1}, Marco Stangalini\altaffilmark{3,1}, Fabrizio Giorgi\altaffilmark{1}}\email{mariarita.murabito@inaf.it}

\altaffiltext{1}{INAF - Osservatorio Astronomico di Roma, Via Frascati,33, Roma Italy}
\altaffiltext{2}{Dipartimento di Fisica e Astronomia - Sezione Astrofisica, Universit\`{a} degli Studi di Catania, Via S. Sofia 78, 95123 Catania, Italy}
\altaffiltext{3}{ASI, Agenzia Spaziale Italiana, Via del Politecnico snc, 00133 Rome, Italy}
\shorttitle{}
\shortauthors{Murabito et al.}

\begin{abstract}

Penumbral transient brightening events have been attributed to magnetic reconnection episodes occurring in the low corona. We investigated the trigger mechanism of these events in active region NOAA 12546 by using multi-wavelength observations obtained with the Interferometric Bidimensional Spectrometer (IBIS), by the \textit{Solar Dynamics Observatory} (SDO), the  \textit{Interface Region Imaging Spectrograph} (IRIS), and the \textit{Hinode} satellites. We focused on the evolution of an area of the penumbra adjacent to two small-scale emerging flux regions (EFRs), which manifested three brightening events detected from the chromosphere to the corona. Two of these events correspond to B-class flares. The same region showed short-lived moving magnetic features (MMFs) that streamed out from the penumbra. In the photosphere, the EFRs led to small-scale penumbral changes associated with a counter-Evershed flow and to a reconfiguration of the magnetic fields in the moat. The brightening events had one of the footpoints embedded in the penumbra and seemed to result from the distinctive interplay between the pre-existing penumbral fields, MMFs, and the EFRs. The \textit{IRIS} spectra measured therein reveal enhanced temperature and asymmetries in spectral lines, suggestive of event triggering at different height in the atmosphere. Specifically, the blue asymmetry noted in \ion{C}{2} and \ion{Mg}{2} h\&k lines suggests the occurrence of chromospheric evaporation at the footpoint located in the penumbra as a consequence of magnetic reconnection process at higher atmospheric heigths. 

\end{abstract}


\section{Introduction}
\label{sec:introduction}

Present-day high resolution observations reveal the dynamic fine scale structure of sunspots created by magnetoconvective interactions \citep{Borrero2011,Rempel2011}. Small-scale features in sunspot (umbral and penumbra) vary in space and time due to a number of different processes. 
These include oscillations, waves, jets of plasma, and magnetic reconnection. That is believed to lead to flaring events. These latter phenomena, involve plasma heating, particle acceleration, and the release of electromagnetic energy from X-rays to radio wavelengths \citep{Shibata2011,Benz2017}.


Smaller-scale energy release phenomena detected over and near sunspots penumbrae have been related to new magnetic elements in emerging flux regions \citep[EFRs, e.g.][]{Guglielmino12,Cheung14} and to magnetic features moving away from the sunspot toward the boundary of the moat region \citep[MMFs, see, e.g.][and references therein]{Criscuoli2012,Li2019}, which can cancel with pre-existing magnetic fields.

In this regard, \citet{Kano2010} analyzed microflares around a well-developed sunspot, by using \textit{Hinode} \citep[][]{Kosugi2007} satellite data from the X-ray Telescope \citep[XRT,][]{Golub2007} and the Narrowband Filter Imager mounted on the Solar Optical Telescope \citep{Tsuneta2008}. They found that half of the observed microflares were caused by magnetic flux cancellation (\textit{encounters of opposite polarities} in their paper) and when the latter is the main responsible, the microflare has one of the X-ray loop connecting the penumbra to the opposite-polarity patch of an EFR or a MMF embedded in the moat.      

Recently, \citet{Bai2016} reported on a penumbral transient brightening observed with state-of-the-art instruments installed in the \textit{Interface Region Imaging Spectrograph} \citep[IRIS,][]{Depon2014} satellite and 1.6~m \textit{New Solar Telescope} \citep[][]{Cao2010,Goode2010} at the Big Bear Solar Observatory. This penumbral brightening, whose estimated thermal energy was in the range of nanoflares {($10^{22}-10^{25}$~erg)}, that displayed  signatures from the chromosphere to the corona. The same authors found that a MMF had appeared close to the penumbral boundary and at the same location of one of the footpoints associated with the observed brightening. \citet{Bai2016} attributed the triggering mechanism of the analyzed event to magnetic reconnection occurring in the low corona and explained the brightening seen in transition region (TR) and chromosphere as due to the local plasma heated up by downward propagating accelerating particles and thermal conduction.
However, due to very weak signals of the observations analyzed, those authors could not find any evidence of the chromospheric evaporation that is expected to follow the heating of the chromospheric plasma from reconnection processes in the low corona. This evaporation has been reported from analysis of larger flares \citep{Tian2014,Cauzzi15}, as well as in micro- \citep{Chen2010} and nano-flares \citep{Testa2014} observed outside penumbrae. Since the trigger mechanism of the studied brightening could not be clearly identified by their analysis, \citet{Bai2016} solicited more research on the formation process of  brightening events observed in penumbral regions.

Indeed, it is well-known that the magnetic reconnection can occur on any spatial or temporal scale in the solar atmosphere \citep{Priest2000}. Both the current high-resolution observations and magneto-hydrodynamical (MHD) numerical simulations indicate that interactions of EFRs with pre-existing ambient magnetic fields \citep{Cheung14,Archontis2012,Schmieder2014} play a prominent role in models of large- \citep[e.g.,][and references therein]{Louis2015} and small-scale eruptive events \citep[e.g.,][and references therein]{Guglielmino2010,Guglielmino2018,Guglielmino2019}. In particular, it is expected that the magnetic reconnection can occur at different atmospheric heights depending on the 
overlying and/or background magnetic fields and on the strength and size of the EFR \citep{Archontis2004,Mactaggart2015}. Furthermore \citet{Galsgaard2005, Galsgaard2007} demonstrated that the magnetic reconnection strongly depends on the relative orientation between the magnetic field components of the emerging flux and the pre-existing magnetic field. In fact, only when the two flux systems have almost antiparallel orientation the magnetic reconnection is efficient.

In this study, we analyzed multi-wavelength observations of the active region (AR) NOAA 12546 obtained with state-of-the-art instruments to further investigate the trigger mechanism of penumbral brightening events. In particular, we studied the evolution of the penumbral area adjacent to two small-scale EFRs emerged in the AR on 2016 May~20. The analyzed region showed three brightening events detected from the chromosphere to the corona.

The paper is organized as follows: in the next Section we describe the observations and the data processing applied. In Section 3 we feature our analysis and results, which are summarized and discussed in Section 4. Section 5 presents our conclusions.


\section{Data and methods}
\subsection{Observations}

 We analyzed high spatial, spectral, and temporal resolution data acquired by the Interferometric Bidimensional Spectrometer \citep[IBIS,][]{Cav06} at the \textit{Dunn Solar Telescope} of the National Solar Observatory, the Helioseismic and Magnetic Imager \citep[HMI,][]{Sch12} and Atmospheric Imaging Assembly \citep[AIA,][]{Lemen2012} instruments on board the \textit{Solar Dynamics Observatory} \citep[SDO,][]{Pesnell2012} satellite, and the \textit{IRIS} and \textit{Hinode}/XRT space-borne telescopes.

IBIS observations were carried out on 2016 May~20, when the AR was characterized by a $\beta$ magnetic configuration.
The dataset consists of full-Stokes measurements taken along the \ion{Fe}{1} 617.30~nm and \ion{Ca}{2} 854.20~nm lines, each line sampled at 21 spectral positions over a Field-of-View (FoV) of about $40\arcsec \times 90\arcsec$. The data were acquired with a spectral sampling of 20~m\AA{} and 60~m\AA, and a spatial resolution of 0\farcs16 and 0\farcs23 for the \ion{Fe}{1}  and \ion{Ca}{2} measurements, respectively, with a cadence of 48~s under excellent seeing conditions that lasted about 180 minutes (318 line scans). The data were processed with the methods described by, e.g., \citet{Ermolli2017}. 
The same dataset was also analyzed by \citet{Stanga2018} and \citet{Murabito19}, which we refer to the reader for further details.

\textit{SDO} data comprise the Space-weather HMI Active Region Patches \citep[SHARPs,][]{Bobra14} continuum filtergrams, magnetograms and Dopplergrams derived from the HMI measurements at the \ion{Fe}{1} 617.3~nm line, performed  with a resolution of $1\arcsec$ and a cadence of 12 minutes from 2016 May~19 at 08:00~UT to May~21 at 08:00~UT. 
Furthermore, we considered AIA filtergrams taken at the 1600, 304, 171, 335, and 131 \AA{} bands (hereafter referred to as A1600, A304, A171, A335, A131, respectively), with a pixel scale of about 0\farcs6 and a cadence of 12~s and 24~s for the EUV and UV channels, respectively. 

The \textit{IRIS} dataset was acquired at the time of the IBIS measurements on 2016 May~20 from 13:17~UT to 16:30~UT. It consists of a sit-and-stare scan (OBS362011063) taken with a cadence of 20~s at the \ion{C}{2} 1334.53~\AA, \ion{Si}{4} 1402.77~\AA, and \ion{Mg}{2} h\&k 2796.35 and 2803.53~\AA{} lines. Simultaneous slit-jaw filtergrams (SJIs) were acquired in the passbands of the \ion{Si}{4} 1400~\AA{} line and \ion{Mg}{2}~k line wing  (hereafter referred to as I1400 and I2832), with a cadence of 20~s and 97~s, respectively, covering a FoV of $120\arcsec \times 119\arcsec$. The I1400 and I2832 data sample plasma at $\mathrm{T} = 65,000$~K and $\mathrm{T} = 6,000 - 10,000$~K, respectively.  

Finally, we analyzed
 \textit{Hinode}/XRT images taken through the Al polyimide and the Be thin filters, whose temperature response ranges $6< \log \mathrm{T}<7.5$ for the former, and has a peak at $\sim \log \mathrm{T}=7$ for the latter \citep{Golub2007}. The two sets of images were acquired with a cadence of 60~s and varying exposure time. These data provide measurements over a FoV of $384\arcsec\times384\arcsec$, with a pixel size of 1\farcs03. Note that XRT observations were taken simultaneously to IRIS data with some gaps. 


\begin{figure}
\centering
	\includegraphics[scale=1.,clip, trim=180 330 160 310]{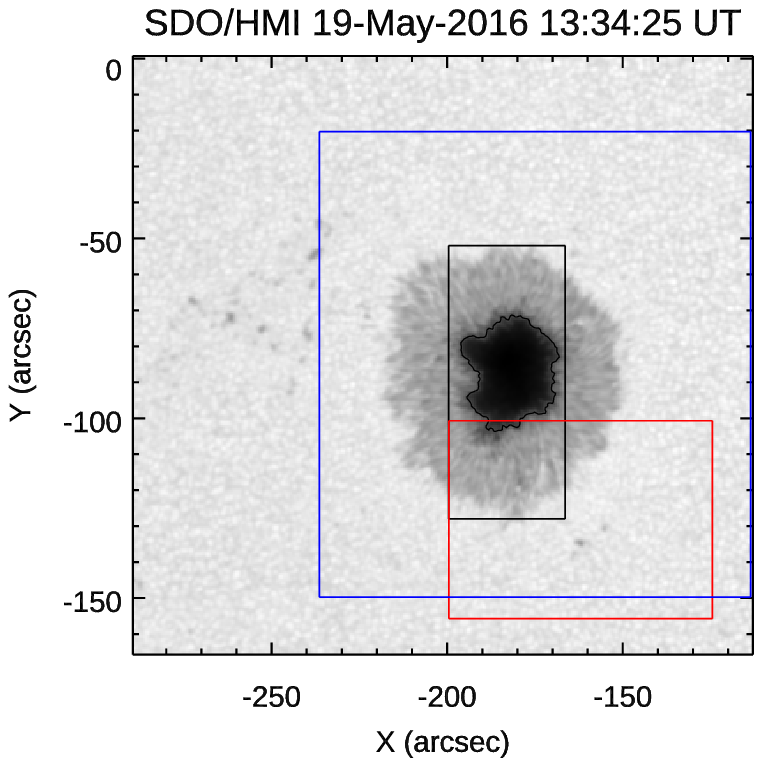}\\
	\includegraphics[scale=1.,clip, trim=180 305 160 330]{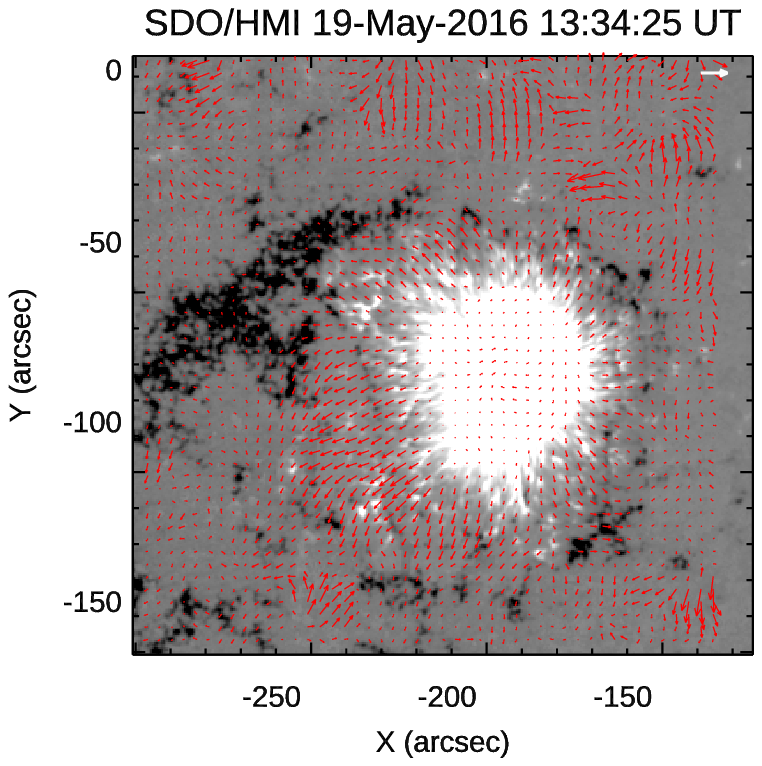}\\
	\caption{Sub-array extracted from the full-disk continuum filtergram (top panel) and LOS magnetogram (bottom panel) taken by \textit{SDO}/HMI on May~19 at 13:34~UT and centered on the studied AR. Values of the LOS component of the magnetic field are saturated at $\pm 500$~G. The black and blue boxes indicate the FoV of the IBIS and \textit{IRIS} observations analysed in our study, respectively, while the red box shows the ROI where the studied brightening events occur. Here and in the following figures, solar North is to the top, and West is to the right. The horizontal velocity estimated with FLCT is overplotted on the LOS magnetogram. The white reference arrow indicates a horizontal velocity of plasma of $1$~km~s$^{-1}$. See Sections~2 and~3 for more details.}
	\label{Figure:1}
\end{figure}  
   
\begin{figure}
	\includegraphics[scale=0.38,clip, trim=105 50 50 200]{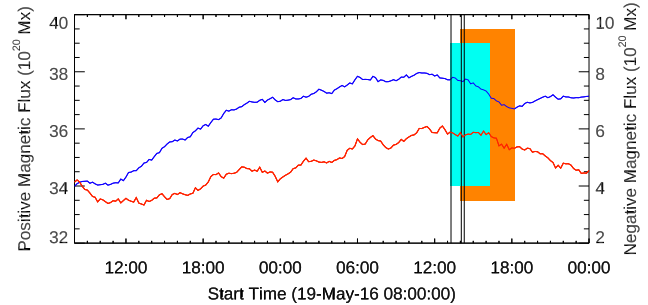}\\
	\caption{Evolution of positive (red) and negative (blue) magnetic flux in the ROI (marked with the red box in Figure~\ref{Figure:1}) as derived from \textit{SDO}/HMI measurements. The time interval of the \textit{IRIS} sit-and-stare scan and of the IBIS observations is indicated with blue and orange strips, respectively. The vertical lines indicate the three brightening events discussed in the text.}
	\label{Figure:2}
\end{figure}

\subsection{Data processing}

We co-aligned IBIS, \textit{SDO}/HMI and \textit{SDO}/AIA observations by applying cross-correlation techniques on co-spatial FoVs that were extracted from the three datasets. In particular, we used the first IBIS spectral image taken at the \ion{Fe}{1} 617.3~nm line continuum on 2016 May~20 at 13:53~UT as a reference and the corresponding, neighboring-in-time \textit{SDO}/HMI continuum image and \textit{SDO}/AIA A1600 filtergram. We employed IDL \textit{SolarSoft} mapping routines to account for the different pixel size of the data. Since the \textit{SDO}/AIA data are aligned between them we employed the \textit{SDO}/AIA A1600 image aligned to the IBIS data as a reference for the remaining \textit{SDO}/AIA channels. Then, we aligned the \textit{IRIS} data to the other measurements, by using the \textit{IRIS} SJIs I2832 filtergrams and the closest in time \textit{SDO}/HMI continuum images. Finally, we aligned the XRT images using the \textit{SDO}/AIA A131 filtergrams as an anchor channel. The precision of our data alignement is comparable to the pixel size of the \textit{SDO} observations, about 0\farcs6,  which is accurate enough for the analysis presented in the following.

To get quantitative estimates of the physical parameters in the analyzed region, we inverted the IBIS data by using the non-LTE inversion code NICOLE \citep{Nicole2015}. In brief, we assumed five equidistant nodes for temperature, three nodes for each component of the vector magnetic field (B$_x$, B$_y$, and B$_z$), two nodes for the line-of sight (LOS) velocity, and one node for both the microturbulence and macroturbulence.  We performed the inversions in two cycles, by assuming as the starting guess model a modified FALC atmosphere \citep{FAL-C} with a constant value of 1.5~kG for B$_z$. We refer the reader to the paper by \citet{Murabito19} for further details.

In order to study the horizontal proper motions and estimate the velocity of the photospheric plasma, we applied the Fourier Local Correlation Tracking technique \citep[FLCT,][]{Fisher2008} to the available \textit{SDO}/HMI LOS magnetograms. We set the FWHM of the Gaussian tracking window to 15 pixels, in order to follow the collective motions of the magnetic structures, and made a temporal integration over a time interval of 12 minutes. 

In addition, we got pixelwise estimates of the LOS velocity of the photospheric plasma by computing the Doppler shift of the line core of the IBIS \ion{Fe}{1} data with respect to the average position of the line core in a quiet Sun region. We computed the line core position by fitting the Stokes~I \ion{Fe}{1} measurements with a linear background and a Gaussian function. 


\section{Analysis and Results}
\label{sec:result}

\begin{figure*}
	\centering
	\includegraphics[scale=0.68, clip, trim=100 7 50 0]{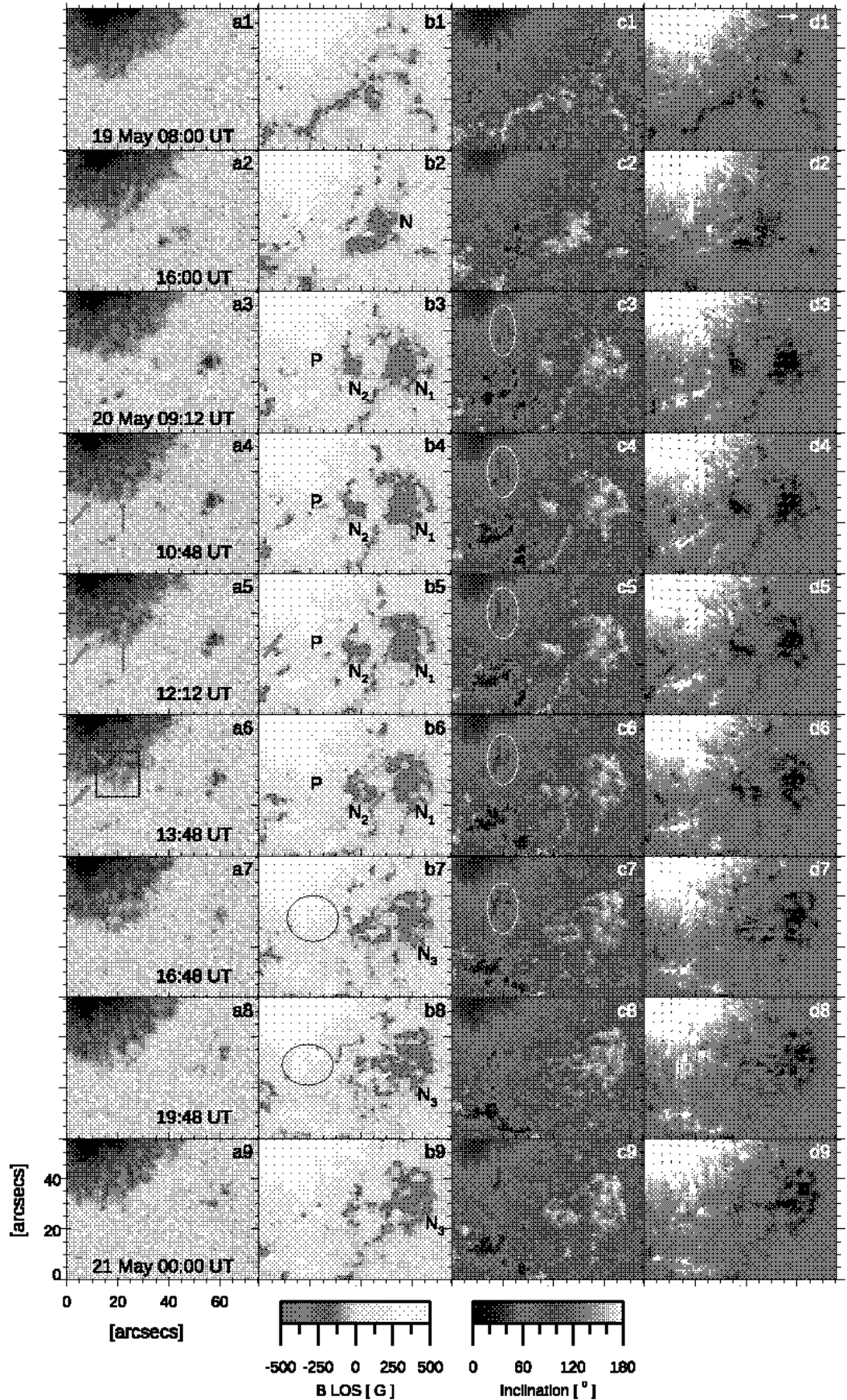}
	\caption{Evolution of the ROI as inferred from \textit{SDO}/HMI data. From left to right, top to bottom: maps of (a) the continuum intensity, (b) LOS magnetic field strength, (c) magnetic field inclination, (d) and plasma horizontal flow velocity for representative times from 2016 May~19 08:00~UT to May~21 00:00~UT that manifest the main changes occurred in the area. Labels, arrows, circles, and ovals overplotted on the maps point to features described in Section~3.1. The blue box in the continuum map taken on May 20 13:48 UT indicates the IBIS sub-FoV shown in the Figure~\ref{Figure:5}. The white arrow overplotted on the top-right panel represents a horizontal velocity of $1$~km~s$^{-1}$. 
	}
	\label{Figure:4}
\end{figure*}

\begin{figure}
	\centering
	\includegraphics[scale=0.63, clip, trim=70 60 0 0]{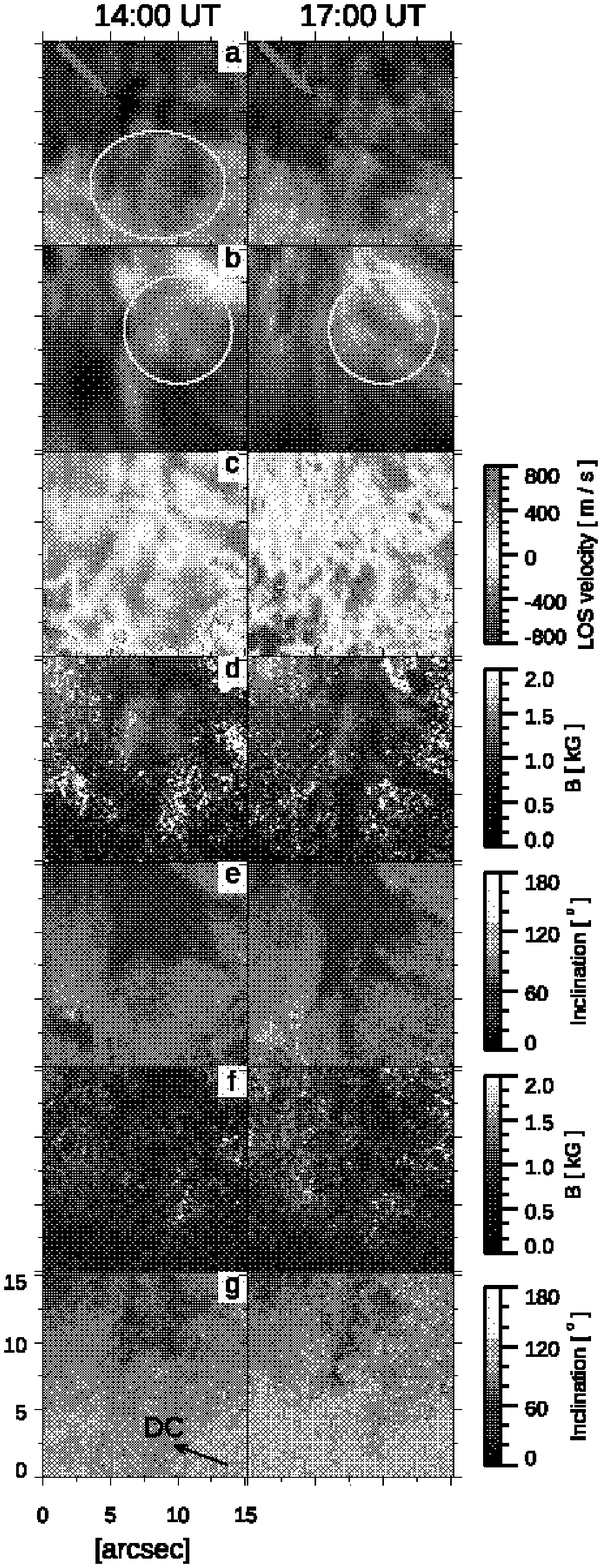}
	\caption{Examples of the IBIS observations analyzed in our study. From top to bottom, the various panels display: maps of (a) the \ion{Fe}{1} continuum intensity,  (b) \ion{Ca}{2} line core intensity, (c) LOS velocity derived from the Doppler shift of the \ion{Fe}{1} line center, (d) total magnetic field strength and (e) magnetic field inclination at photospheric height $\log \tau=-1$, (f) total magnetic field strength and (g) magnetic field inclination at the chromospheric height $\log \tau=-4.6$. Downflows and upflows correspond to positive and negative velocity values, respectively. Arrows, ovals, and circles indicate features described in the text. The black arrow in panel (g) points to the disk center. \\
    (An animation of the white light images and \ion{Ca}{2} line core measurements from all IBIS observations is available in the online material.) 
    }
	\label{Figure:5}
\end{figure}

From 2016 May 15 to May 25, ground-based and space-borne solar telescopes detected the on-disk passage of AR NOAA 12546 (AR hereafter), which emerged in the southern solar hemisphere close to the equator. This AR consisted  of a leading, large and almost circular sunspot and a trailing, extended plage region. The sunspot, which was one among the largest such structures observed over the solar cycle 24, kept its regular shape nearly unchanged during the disk passage, while the plage region evolved significantly. 
The studied sunspot also displayed a large number of MMFs, mostly of type II, i.e., unipolar features with the same polarity of spot \citep[e.g.,][and references therein]{Zuccarello2009}, seen to move almost radially away from the sunspot structure toward the boundary of the moat region. 

Both the \textit{SDO}/HMI and IBIS photospheric observations show that the penumbra of AR was formed by bright filaments and dark spines  nearly radially aligned over most of the sunspot structure \citep{Murabito19}, except for an area close to the EFRs observed on 2016 May 20. 

\subsection{Magnetic environment in the photosphere}

Figure~\ref{Figure:1} shows the region analyzed in our study as deduced from \textit{SDO}/HMI continuum observation performed on 2016 May~19 at 13:34~UT and the region of interest (ROI) discussed in the following and shown in Figures~\ref{Figure:4}, \ref{Figure:9}, and~\ref{Figure:12}. The bottom panel of Figure~\ref{Figure:1} displays the LOS magnetogram with over-plotted horizontal velocity.

Starting from 2016 May~19 12:00~UT the ROI showed in about 24 hours an increase of the LOS magnetic flux 
of about $2 - 4 \times 10^{20}$~Mx for the positive and negative polarity, and a flux decrease afterwards. This flux variation, which is reported in Figure~\ref{Figure:2}, followed the evolution of the region summarized in the panels of Figure~\ref{Figure:4}, which show maps from SDO/HMI observations at 9 representative times from 2016 May~19 08:00~UT to May~21 00:00~UT.  
 
On 2016 May~19 08:00~UT the ROI only included the southern sector of the penumbra and moat area nearby (Figure~\ref{Figure:4}, first row). A few hours later, at 16:00~UT, negative polarity patches (labelled N in Figure~\ref{Figure:4}, panel b2) from new flux of an EFR emerged at X = [20\arcsec, 50\arcsec], Y = [15\arcsec, 40\arcsec], had already merged with pre-existing network field forming a small pore and a few fainter structures (labelled N$_1$ and N$_2$ in Figure~\ref{Figure:4}, panel b3). These structures also received the negative polarity flux from a second EFR appeared 
on 2016 May 20 01:00~UT. Thereafter, positive polarity patches streaming out from the penumbra and from the second EFR merged to form small scale structures in the south-east section of the ROI (at X = [5\arcsec, 30\arcsec], Y = [10\arcsec, 25\arcsec]) that later on, e.g., at 13:48~UT, were only merely discernible. 

During the emergence of the second EFR, a bright lane (marked with red arrows in Figure~\ref{Figure:4}, panels a4-a5) and a few dark patches (marked with the orange arrow in Figure~\ref{Figure:4}, panel a6) appeared in the penumbra near the EFR, by creating a gap in the structure (see green arrows in Figure~\ref{Figure:4}, panels a4 -a6) that separated filaments with different orientation.

\begin{figure}
	\centering
	\includegraphics[scale=0.4, clip, trim=60 80 50 20]{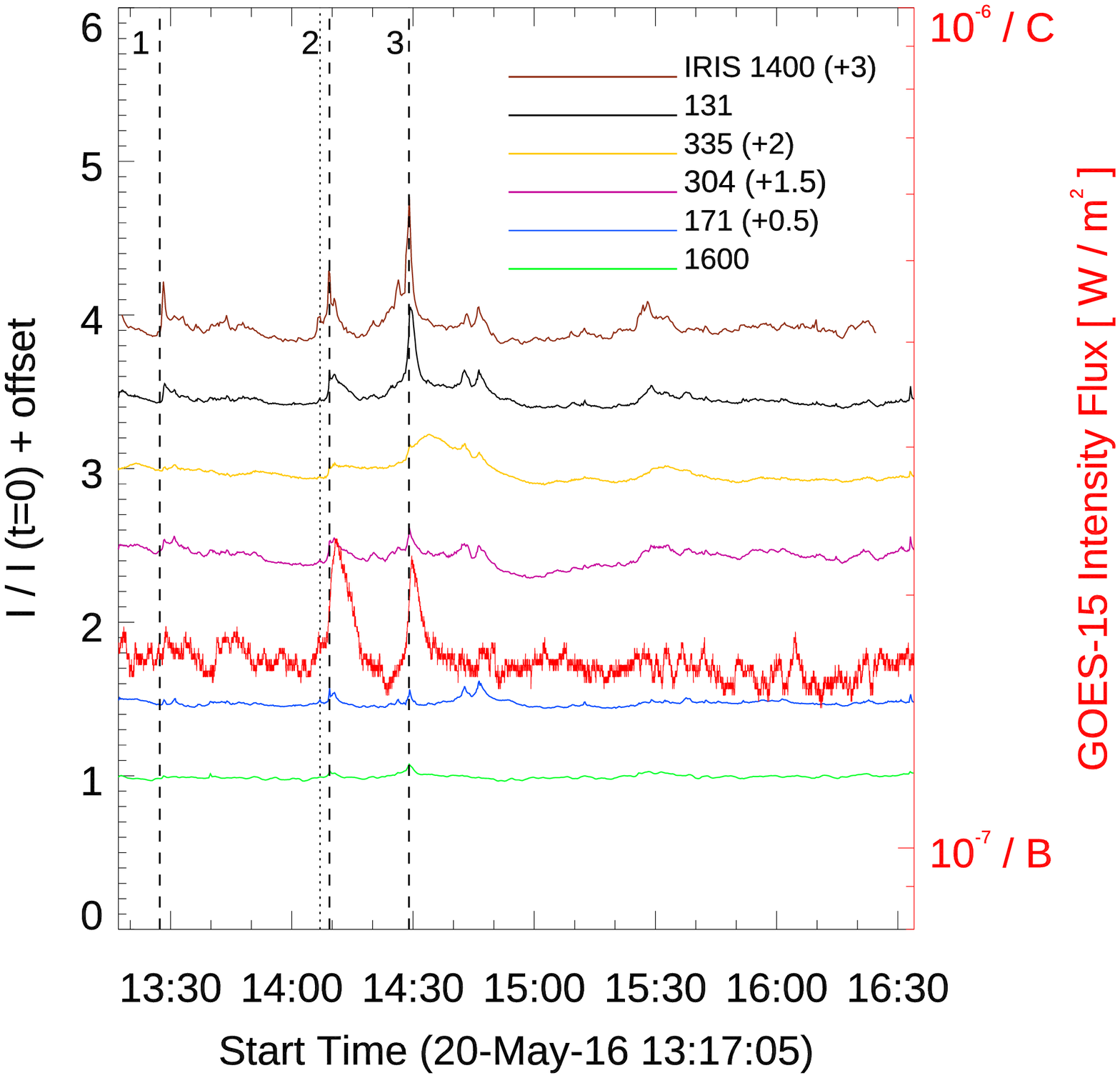}\\
		\includegraphics[scale=0.4, clip, trim=60 0 50 260]{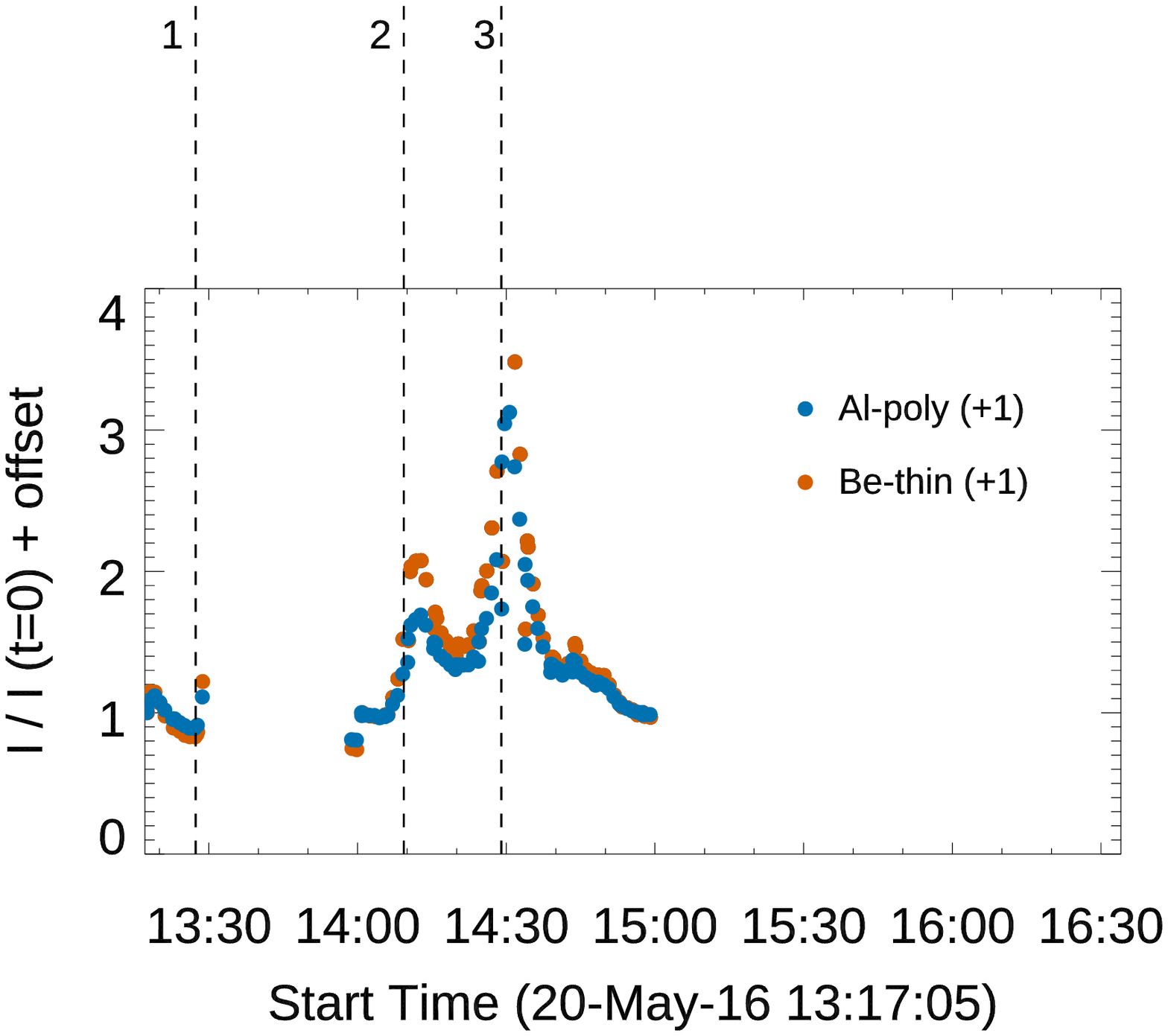}\\
	\caption{Top panel: Light curves computed from the \textit{SDO}/AIA observations and \textit{IRIS}/SJI 1400~\AA{} data obtained on 2016 May~20 from 13:17~UT to 16:30~UT. The red line represents the X-ray flux measured by \textit{GOES}-15 in the $1 - 8$~\AA{} channel. Note the offsets of the light curves along the \textit{y} direction, to enhance their visibility. The vertical dotted line indicates the starting time of IBIS observations. The three dashed green lines mark the occurrence of the three brightening events \textit{E1}, \textit{E2}, and \textit{E3}, which are described in Sections~3.2 and 3.3 and shown in Figures~\ref{Figure:9}, \ref{Figure:12}, and~\ref{Figure:13}. Bottom panel: Blue and orange dots represent X-ray light curves obtained from \textit{HINODE}/XRT data through the Al pol and Be thin filters, respectively. }
		\label{Figure:8}
\end{figure}

\begin{figure*}
	\centering
	\includegraphics[scale=0.8, clip, trim=0 575 0 65]{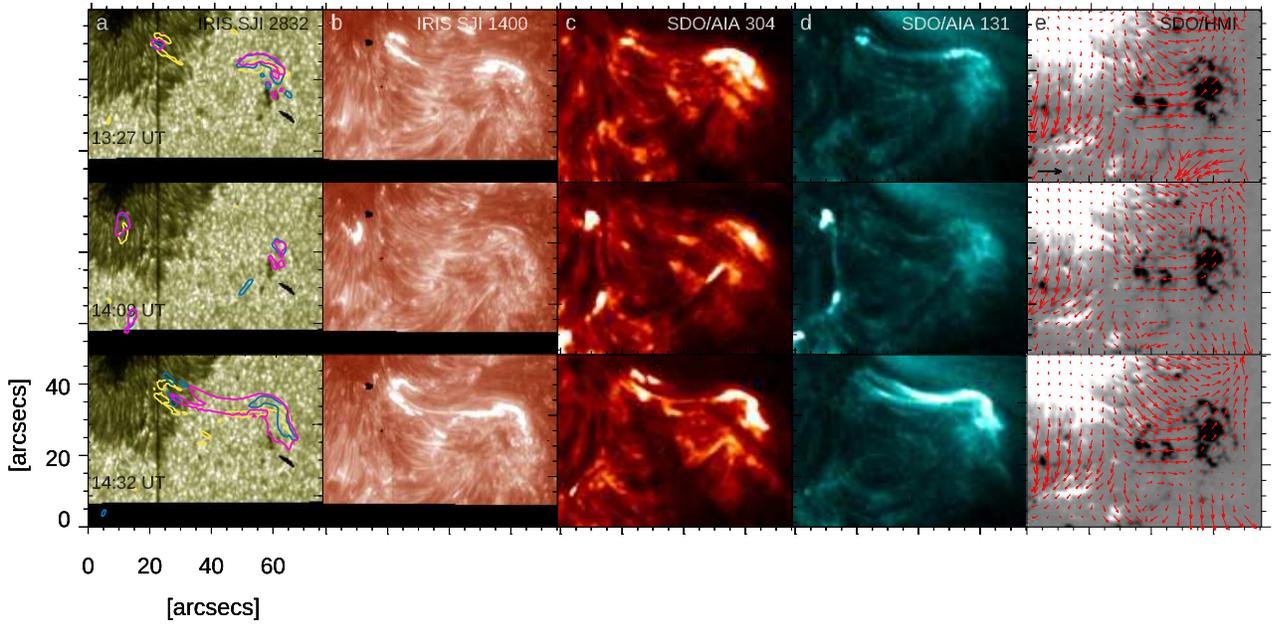}
	\caption{Examples of the \textit{SDO} and \textit{IRIS} data analyzed in our study, relevant to the ROI. From left to right: (a) \textit{IRIS} I2832, (b) \textit{IRIS} I1400, (c) \textit{SDO} A304, and (d) \textit{SDO} A131 filtergrams. Column (e) displays the maps of the LOS magnetic field from \textit{SDO}/HMI data with overplotted the horizontal velocity field derived from the same observations. The light curves in Figure~\ref{Figure:8} were computed over the FoV shown in these panels. The colored contours overplotted on the \textit{IRIS} I2832 data represent the footpoints and regions of maximum brightening in the \textit{IRIS} I1400 (green), \textit{SDO}/AIA A131 (red), and A304 (blue) measurements, respectively. The black arrow in panel (e) represents a horizontal velocity of $1 \mathrm{km\, s}^{-1}$. 
	}
	\label{Figure:9}
\end{figure*}

\begin{figure*}
	\centering
	\includegraphics[scale=0.9, clip, trim=20 400 50 55]{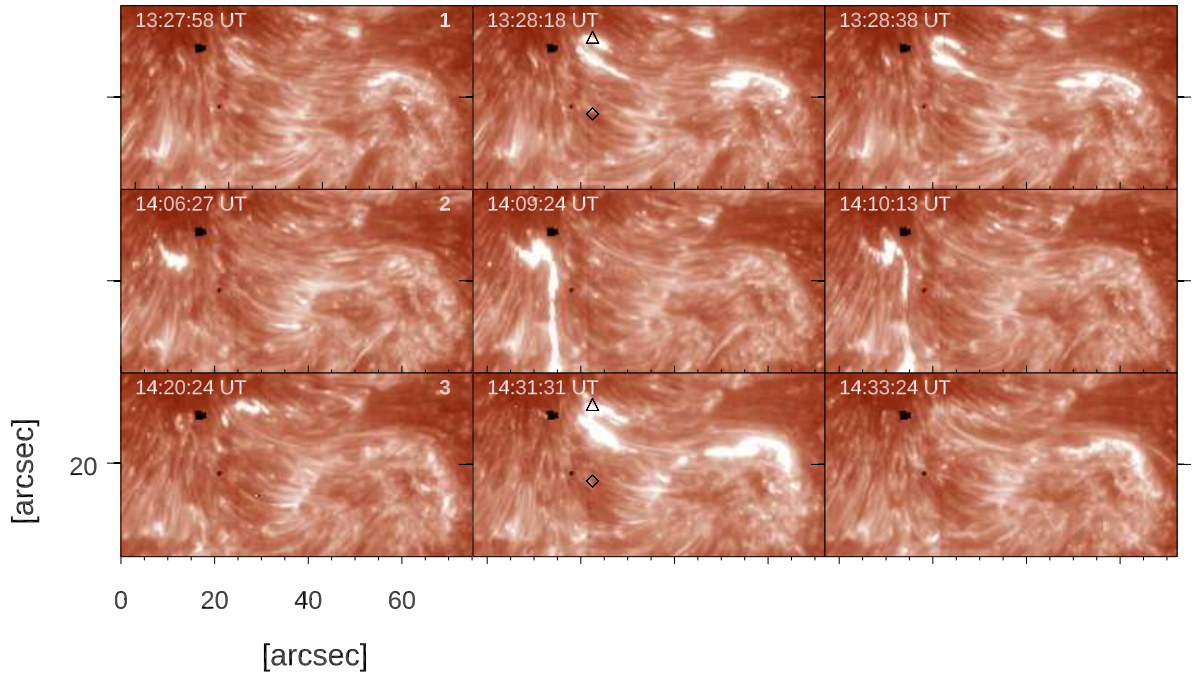}
	\caption{Sequences of \textit{IRIS} I1400 images showing the evolution of the penumbral brightening events labelled as \textit{E1}, \textit{E2}, and \textit{E3} in Figure~\ref{Figure:8} and described in the Sections~3.2 and~3.3. From left to right, top to bottom: observations taken before, at, and after the maximum brightening of the \textit{E1}, \textit{E2}, and \textit{E3} events, respectively.}
		\label{Figure:12}
\end{figure*}

\begin{figure*}
	\centering
	\includegraphics[scale=1.4, clip, trim=20 500 250 50]{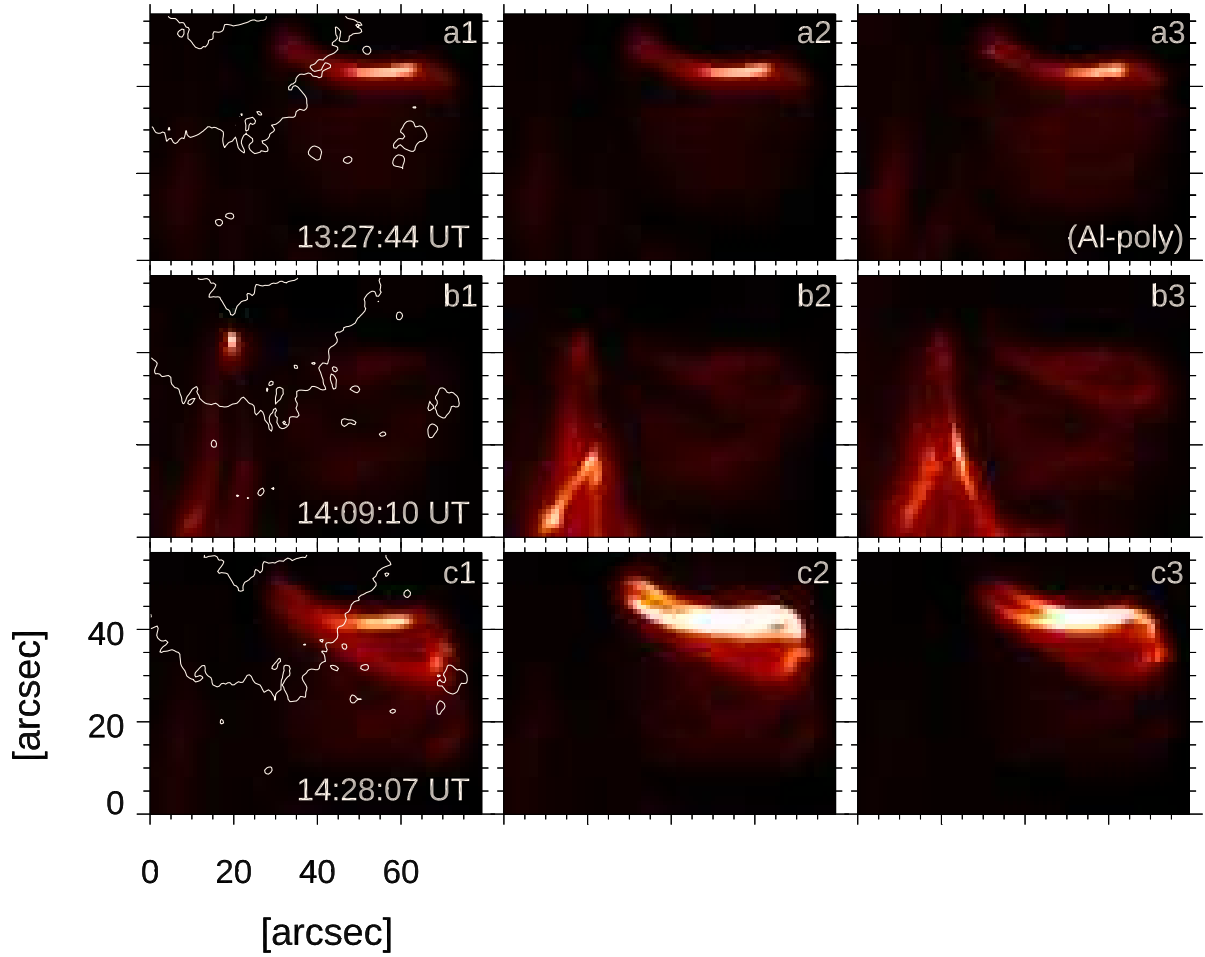}
	\caption{X-ray intensity through the \textit{Hinode}/XRT Be thin filter at the time of \textit{E1} (panels a1-a3), \textit{E2} (panels b1-b3), and \textit{E3} (panels c1-c3) events. The white contours indicate the umbra and penumbra in the continuum intensity as a reference. } \label{Figure:XRT}
\end{figure*}

\begin{figure*}
	\centering
	\includegraphics[scale=0.5, clip, trim=20 140 50 150]{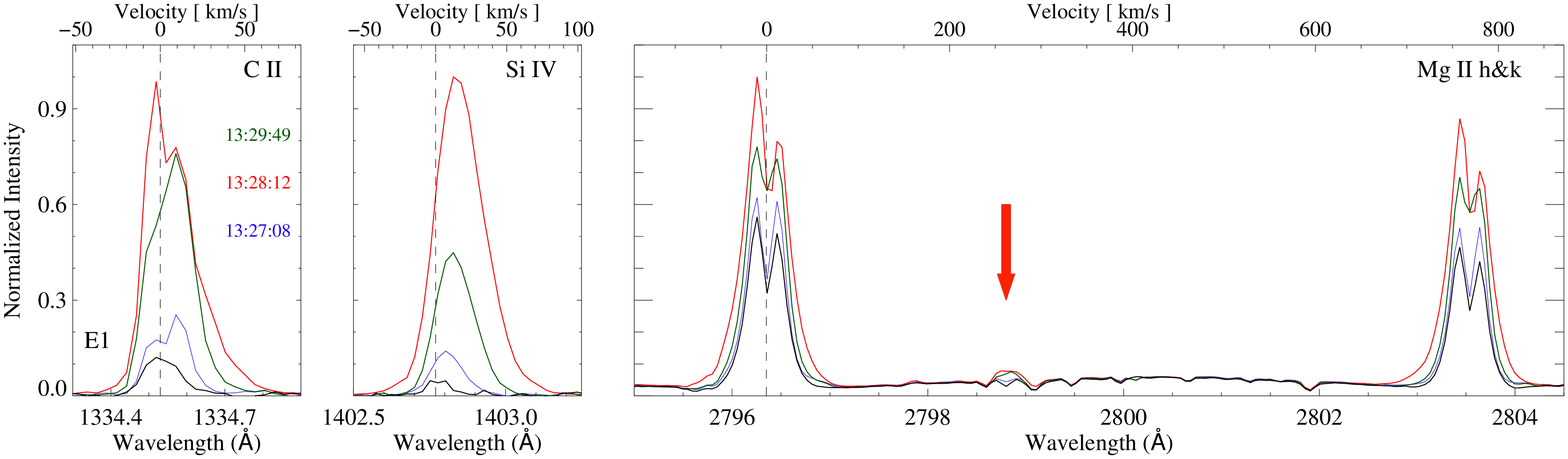}
	\includegraphics[scale=0.5, clip, trim=20 80 50 140]{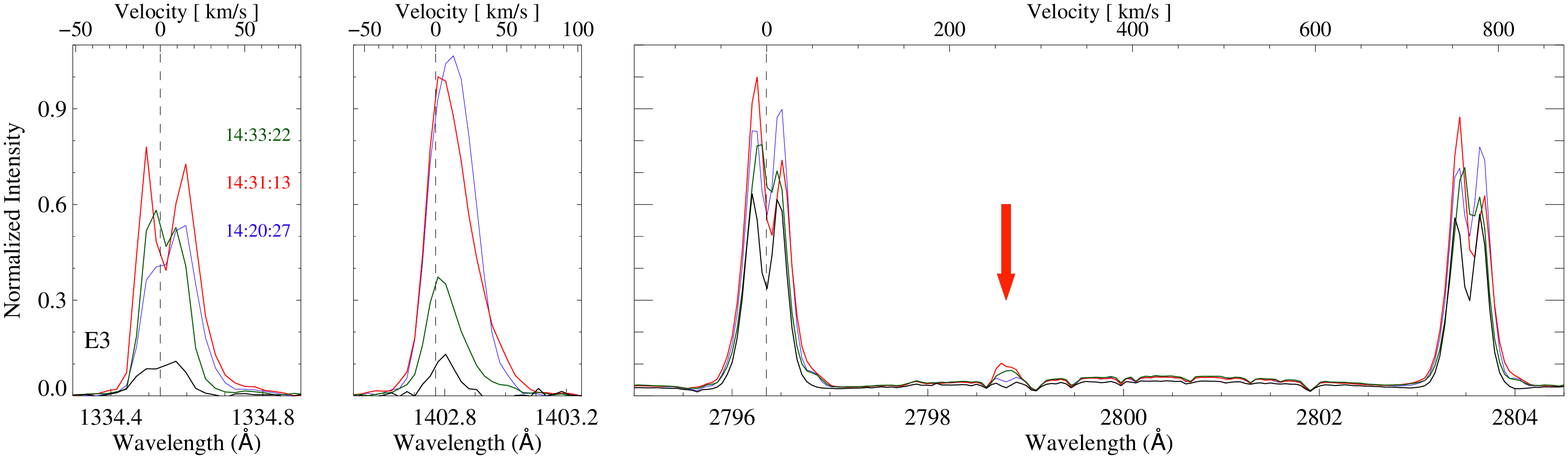}
	\caption{Spectra from \textit{IRIS} observations taken during the \textit{E1} (top panels) and \textit{E3} (bottom panels) penumbral brightening events. From left to right: profiles of the \ion{C}{2} 1334.53~\AA, \ion{Si}{4} 1402.77~\AA, and \ion{Mg}{2}~h\&k lines at three times during each studied event and in a quiet penumbral region considered as a reference. The line profile of the latter region is shown with solid black line. Blue, red, and green lines display the measurements taken before, at, and after the each brightening, respectively. Red arrows indicate the rare weak emission at the \ion{Mg}{2} 2798.8 \AA.
	}
	\label{Figure:13}
\end{figure*}

\begin{figure*}
	\centering
	\includegraphics[scale=1.15,clip, trim=20 405 210 30]{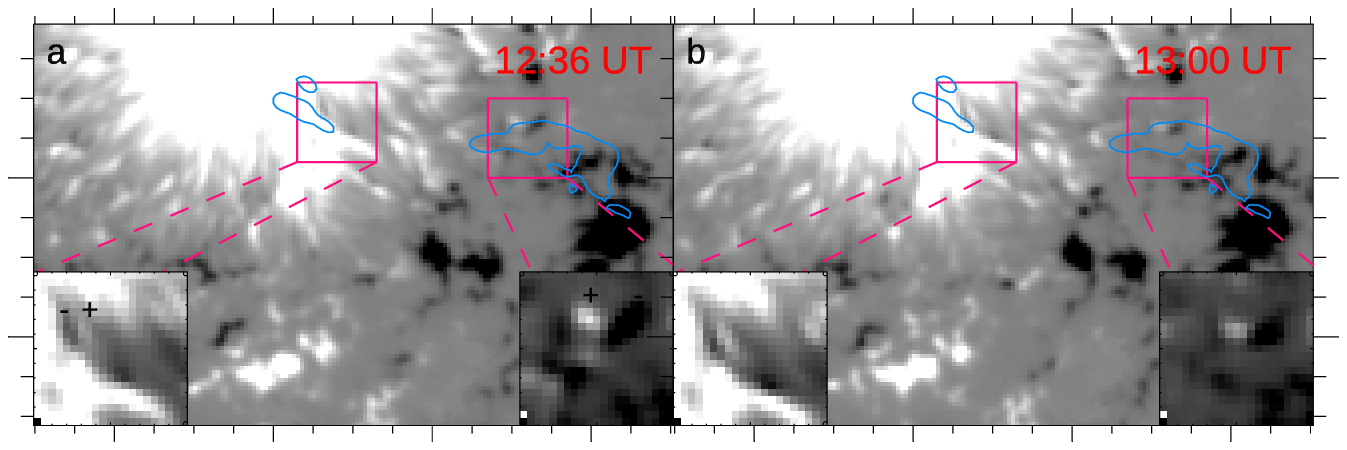}
	\includegraphics[scale=1.15,clip, trim=20 405 210 40]{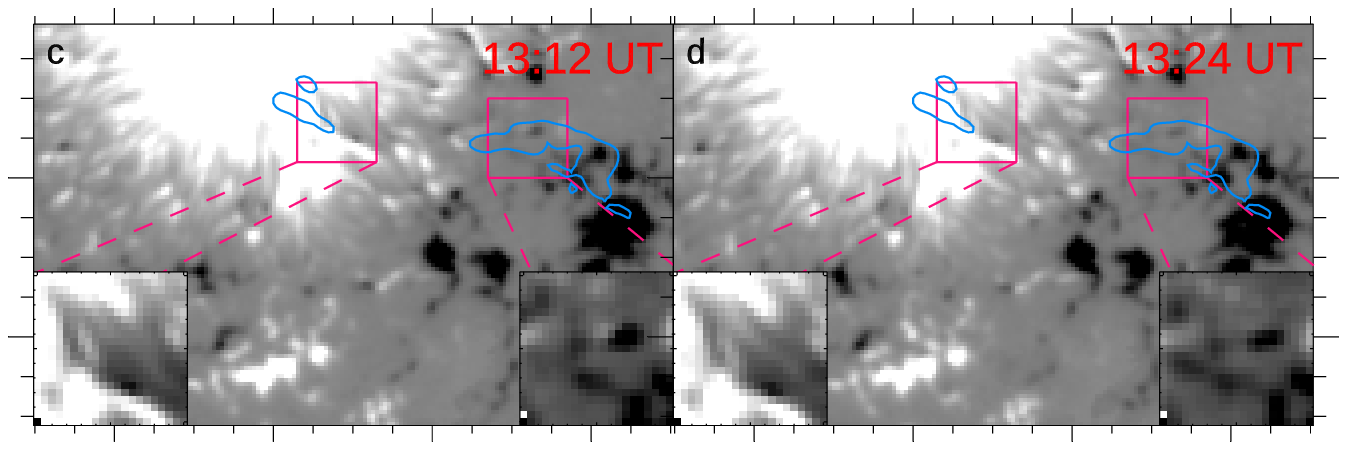}	
	\includegraphics[scale=1.15,clip, trim=20 405 210 40]{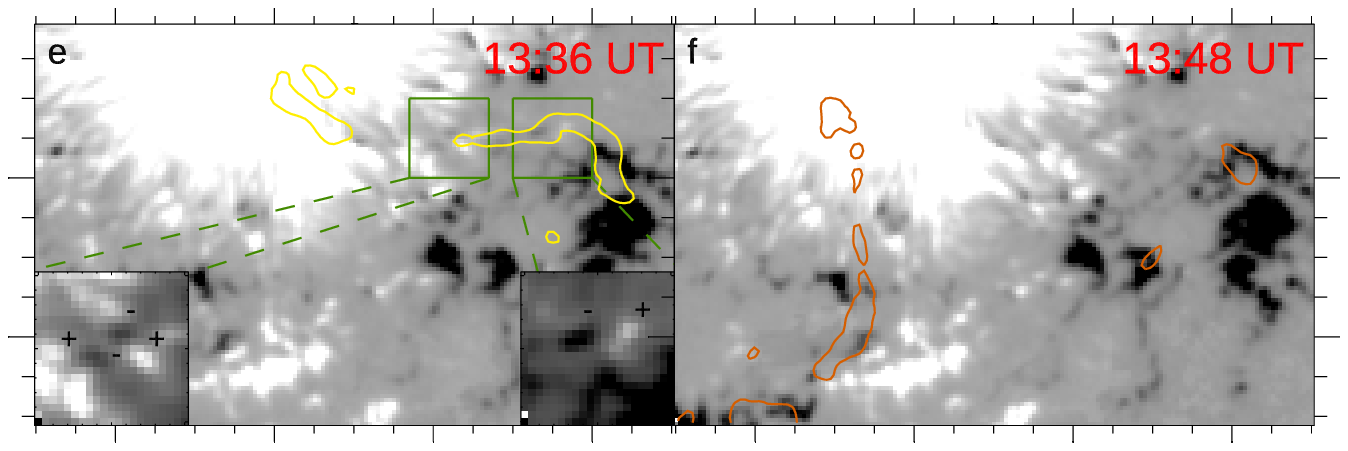}
	\includegraphics[scale=1.15,clip, trim=20 370 210 40]{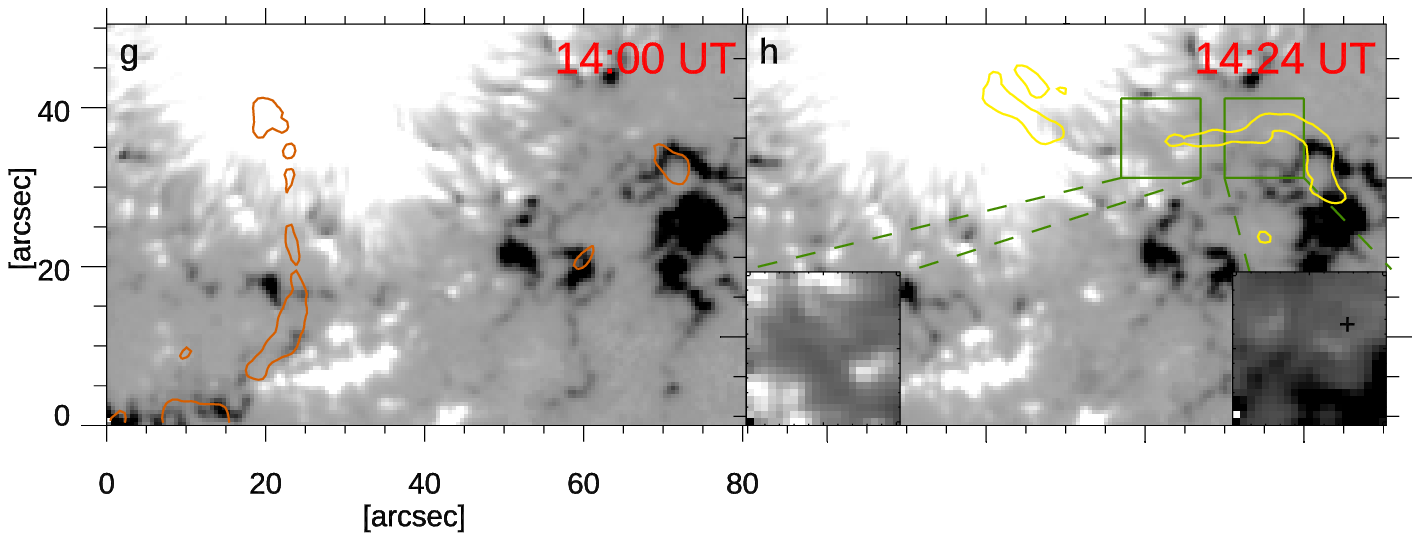}
	\caption{Magnetic evolution of the ROI before the \textit{E1} and \textit{E3} events. The regions in the pink and green squares are shown zoomed in the inserts at the left and right corners of each panel. The light blue, orange and yellow contours mark the intensity pattern at maximum brightening in \textit{SDO}/AIA A304 for \textit{E1}, \textit{E2} and \textit{E3}, respectively. The polarities of the MMFs identified in the inserts are shown by the ``+" and ``-" symbols.
	\label{Figure:15}}
\end{figure*}

Later, the filaments closer to N$_{2}$ (black oval in Figure~\ref{Figure:4}, panels b7-b8) broke away from the penumbra (black oval in Figure~\ref{Figure:4}, panel b8) when the small-scale negative flux patches from the two EFRs and pre-existing fields coalesced and got more aggregated (N$_3$ in Figure~\ref{Figure:4}, panel b7-b9).

The animated  version in Figure~\ref{Figure:4} (panels a and b), available in the online material clearly shows the entire evolution described above and the diverse MMFs discussed below.

The penumbra had a rather homogeneous structure with field inclination of about $60^{\circ} - 90^{\circ}$ (Figure~\ref{Figure:4}, third column), except for a few elongated patches with field inclination in the range $40^{\circ} - 60^{\circ}$ (marked with the white oval in Figure~\ref{Figure:4}, panels c3-c7) in the area that broke away. These features were clearly linked to the positive polarity patches of the EFR. The region also displayed two systematic flow patterns, which are shown in Figure~\ref{Figure:4} (fourth column).
The former, directed towards the first EFR, led to the formation of N$_3$ by the merging of N$_1$ and N$_2$, while the second southern-ward pattern contributed to the evolution of the small-scale positive polarity features. Note that the emergence of both EFRs was characterized by strong shear motions with values of the horizontal velocity of the plasma of about $1 \mathrm{km \, s}^{-1}$.

The IBIS data provide a close-up view of the evolution of the ROI and of the physical properties therein. In particular, we show in Figure~\ref{Figure:5} the analyzed area from the IBIS  observations taken on 2016 May~20 at 14:00~UT (left panels) and 17:00~UT (right panels), along with results from their inversion.
The IBIS data at 14:00~UT indicate that the bright gap in the penumbra (yellow oval) already noticed in the SDO/HMI observations separates filaments with different orientation in the irregularly shaped penumbral sector. 
The dark structures (red arrow), whose length was of about 10\arcsec, decreased in size over time until they were no longer detected, e.g., on 2016 May~20 at 18:00~UT. Noticeably, the plasma LOS velocity in small-scale patches of these structures was opposite than regular Evershed flows nearby (Figure~\ref{Figure:5} panel c). 

Figure~\ref{Figure:5} (panels d-e) also display filaments at the atmospheric height $\log \tau=-1.0$ with average values of the field strength and inclination of about 1.4~kG and about 45$^{\circ}$, respectively, in the area that broke away from the penumbra, compared to the values for the same quantities of 0.9~kG and 80$^{\circ}$ for the homogeneous filaments nearby. 
Later, e.g., at 17:00~UT, the field pattern in the region was unchanged but for the slightly larger extension of the in-homogeneous patches.

\subsection{Signatures from the chromosphere to the corona}

At the chromospheric heights sampled by the IBIS \ion{Ca}{2} 854.2~nm measurements, the ROI showed a bundle of filaments with intense and repeated brightening at various locations, especially in the area of the two EFRs. These small-scale enhancements were observed in the form of either circular or elongated features (see the on line animation of the \ion{Ca}{2} line core images in Figure \ref{Figure:5}). 
Moreover, at $\log \tau=-4.6$ (Figure~\ref{Figure:5}, panels f-g) the ROI exhibited average values of the field strength of about 0.9~kG and field inclination ranging from 60$^{\circ}$ to 100$^{\circ}$, except for some small-scale patches in the inhomogeneous area with field inclination lower than 40$^{\circ}$.

The transients observed in IBIS \ion{Ca}{2} data were also manifested in the evolution of the average intensity measured over the ROI in the \textit{IRIS} I1400,  \textit{SDO}/AIA, and  \textit{Hinode}/XRT observations. Figure~\ref{Figure:8} compares the light curves derived from the above data for the time interval of simultaneous IBIS and IRIS data, along with the variation of the X-ray flux measured by the GOES-15 satellite \citep{Bornmann1996}.

All the light curves shown in Figure~\ref{Figure:8} display a couple of remarkable intensity increases standing out with respect to long-term and smooth variations. In particular, we considered  the three abrupt intensity changes marked with vertical dashed lines in Figure~\ref{Figure:8} and referred in the following to as \textit{E1}, \textit{E2}, and \textit{E3}. Two of these peaks, specifically \textit{E2} and \textit{E3}, had clear counterparts in the variation of the X-ray flux measured by \textit{GOES}-15 (red curve in Figure~\ref{Figure:8}) and correspond to B-class flares. The above intensity increases, which lasted a few minutes for all the considered data, had different amplitudes depending on the event and wavelenght, see, e.g., the peak value for the \textit{E2} and \textit{E3} events.

In an attempt to localize the source regions and to investigate the possible trigger mechanisms of the above events, Figure~\ref{Figure:9} displays \textit{SDO}/HMI, \textit{SDO}/AIA and \textit{IRIS} data taken over the time interval considered in Figure~\ref{Figure:8}. The various panels in Figure~\ref{Figure:9} show brightening events whose footpoints and emission contours are over-plotted on the I2832 observations for ease of comparison. We clearly noticed two homologous small-scale flaring events developed over a sigmoidal region (panels in first and third rows of Figure~\ref{Figure:9}). These events correspond to \textit{E1} occurred at 13:27~UT and \textit{E3} recorded by \textit{GOES}-15 as a B-class flares at 14:32~UT, about one hour later than \textit{E1}. The eastern footpoint of the flaring region was lying in the penumbra, while the other was close to the patch of negative polarity flux N$_{1}$. It is worth noting that the latter footpoint can be easily identified in most of the \textit{SDO}/AIA coronal observations that also show an intensity enhancement in the central part of the sigmoidal flaring area during \textit{E3}. The latter area is only faintly detected in co-temporal chromophere and TR, see e.g., I1400 and A304 maps.

Between the above homologous flaring events, we observed the occurrence of \textit{E2}, a small-scale brightening appeared at about 14:08~UT in the left side of the ROI (panels in the second row of Figure~\ref{Figure:9}) in the form of three elongated small-scale features seen in the A304 and A131 data, and only partly detected in I1400 observations. These bright patches were clearly aligned to penumbral filaments and co-spatial to the small-scale positive flux features described in Section 3.1. 

We further investigated the differences among  \textit{E1}, \textit{E2}, and \textit{E3} by using I1400 observations, some examples of which are shown in Figure~\ref{Figure:12}. The evolution of \textit{E1} was fast (Figure~\ref{Figure:12}, top panels) and with simultaneous signatures seen in the light curves and observations of I1400, A131, and A304. 
The \textit{E2} event lasted about 7 minutes and reached its maximum extension 3 minutes after its start (Figure~\ref{Figure:12}, middle panels) in the form of a bright knot appeared in the penumbra. Three minutes after the first appearance of the bright knot, a thin bright lane connected it to a small bright patch outside the penumbra. This bright lane became brighter with time and reached its maximum extension 4 minutes after its first appearance. \textit{E2} produced the strongest signature in the \textit{GOES}-15 X-ray flux record, but only relatively small changes in the intensity emerging from the low chromosphere to the upper TR compared to the same atmospheric layers during \textit{E3}. The latter transient produced intensity enhancements whose amplitude increased with the atmospheric height. \textit{E3} started at 14:20~UT, lasted about 13 minutes, and reached maximum extension (40\arcsec) 11 minutes after its beginning. The maximum of intensity in the I1400 data occurred 2 minutes later than recorded for all the \textit{SDO}/AIA EUV channels.


In Figure \ref{Figure:XRT} we show the \textit{Hinode}/XRT X-ray intensity maps acquired during the maximum brightening of \textit{E1}, \textit{E2} and \textit{E3} events (second column), as well as one minute before (first column) and after (third column) the three events. These images display the same plasma behaviour as reported in A131 and the strongest coronal brightening observed during \textit{E3} too. In particular, the shape of the \textit{E2} brightening (panels b) indicates the possible connection between opposite polarity parches of the diffuse field and MMF.

\subsection{\textit{IRIS} spectra}

The \textit{IRIS} slit sampled the region 
at X=[41\arcsec].
We show in Figure~\ref{Figure:13} examples of the line profiles measured during \textit{E1} and \textit{E3} at the positions marked by the symbols in Figure~\ref{Figure:12} in both the inhomogenous and regular penumbra. 
From a comparison with Figure~\ref{Figure:9}, we can note that the positions marked with triangles correspond to the eastern footpoints of the flaring loop during both \textit{E1} and \textit{E3}.
 The red curves represent the spectra measured at the time of maximum brightening of each event. 
Unfortunately, there are no similar data available for \textit{E2}.

During the transients, the \ion{C}{2} and \ion{Si}{4} line profiles, which are formed in the TR, were enhanced by a factor of about 10 with respect to the line profiles measured in  the quiet penumbral region (black curves), taken as a reference. 
At the time of maximum brightening \textbf{red curves}, the double-peaked \ion{C}{2} profiles showed blue lobes stronger than the red ones. In addition, 
the red wings of the \ion{C}{2} intensity profiles were broader than the blue wings. 
The \ion{C}{2} profiles showed asymmetric peaks varying in time, specifically the red lobe stronger than the blue one during the initial stages (blue curve) of the brightening, and opposite peak ranking at the time of maximum brightening (red curve). However, during \textit{E3} the \ion{C}{2} intensity profile (Figure~\ref{Figure:13}, bottom-left panel) was less asymmetric than reported for \textit{E1}. 
The \ion{Si}{4} line profiles during the two events were broadened by a factor of about 3 with respect to that measured in the reference regular penumbra,  redshifted, indicating plasma velocities up to about $10 \mathrm{km \, s}^{-1}$. 


Chromospheric \ion{Mg}{2} h\&k lines measurements showed a behaviour similar to \ion{C}{2} spectra, since the double-peaked profiles exhibited blue lobes stronger than the red ones at the peak of the brightening events. Moreover, the emission intensity profiles displayed no significant Doppler shift in all the available data. 
During the transients, the emission of the \ion{Mg}{2} h\&k lines was enhanced by a factor 2 with respect to that measured in the  regular penumbra. During \textit{E1} the profiles were rather symmetric, except for the data taken at maximum brightening (red curve). 
Conversely, before the \textit{E3} transient the profiles displayed red lobes stronger than the blue ones, contrarily to data during the maximum brightening (see blue and red curves in the bottom-right panel of Figure~\ref{Figure:13}). 

Noticeably, the \ion{Mg}{2} intensity profiles measured during both \textit{E1} and \textit{E3} showed emission in the \ion{Mg}{2} 2798.8~\AA{} triplet (marked by the red arrows in Figure~\ref{Figure:13}).

\section{Discussion}
\label{sec:discussion}

We presented our analysis of brightening events observed in the penumbra of AR NOAA 12546. This region  included one among the largest sunspots of the solar cycle 24. The data analyzed in our study consist of multi-wavelength and multi-instrument measurements obtained with state-of-the-art facilities, specifically with the IBIS, \textit{SDO}/HMI, \textit{SDO}/AIA, \textit{IRIS} and \textit{Hinode}/XRT  instruments. The bright filaments and dark spines forming the penumbra were nearly radially aligned over most of the sunspot structure and during its entire disk passage, except for a small area at a given time. We analyzed the phenomena occurring in this small area at representative times. 
We found that:

\begin{enumerate}
\item A first EFR emerged close to the penumbral field, with a flux content of about $3 \times 10^{20}$~Mx.
\item The newly emerged flux coalesced to pre-existing network field and formed a small pore of opposite polarity flux with respect to the one in the umbra. 
\item Another small EFR occurred in the sunspot moat with a flux content of about $5 \times 10^{19}$~Mx. The negative flux patches from this EFR merged with the pore formed from the previous EFR. 
\item During the emergence of the second EFR, a bright gap and some dark patches appeared inside the penumbra, in which the positive polarity of the second EFR was embedded. The bright gap separated filaments with different orientation than the one of the filaments nearby.
\item The above photospheric features were preceded by the formation of elongated patches in the penumbra with magnetic field inclination ranging from 40$^{\circ}$ to 60$^{\circ}$, different with respect to the inclination of the field in the filaments nearby, of about 60$^{\circ}$-90$^{\circ}$. These elongated patches also showed  plasma LOS velocity opposite with respect to the regular Evershed flows nearby.
\item The horizontal plasma flow exhibited two systematic velocity patterns directed from the penumbra towards the first EFR and to the southern side of the ROI. Strong diverging motions with values of the horizontal plasma velocity of about $1 \mathrm{km \, s}^{-1}$ characterized the area of the two EFRs. 
\item Three brightening events occurred during the above evolution of the photospheric magnetic environment, with clear signatures detected from the chromosphere to the corona. All the events had a footpoint located in the penumbra. 
\item Two of these events were recorded by \textit{GOES} as B-class flares. One of these (\textit{E3}) can be also classified as an homologous flare \citep[e.g.,][]{Romano:15}, since it showed the same configuration as a previous flaring event (\textit{E1}). 
The first homologous flaring event lasted less than the second one. It left simultaneous signatures at all the chromospheric, TR, and coronal analyzed data. The second homologous event lasted about 13 minutes. 
The flaring region of \textit{E3} was about 40\arcsec{} wide (29 Mm). It is worth mentioning that another homologous flaring event was seen in the \textit{SDO}/AIA filtergrams after the end of the \textit{IRIS} observations, at 16:39 UT. 
\item \textit{IRIS} spectra relevant to the penumbral footpoints during the two homologous flares reveal enhanced emission and asymmetries in both the \ion{C}{2} and \ion{Mg}{2} h\&k line profiles, and weak emission at the \ion{Mg}{2} 2798.8~\AA{} triplet. In particular, at the time of maximum brightening, the blue lobe profiles are stronger than the red ones.  
\end{enumerate}

Observations of AR NOAA 12546 show a large number of short-living MMFs flowing away from the sunspot, some of them occurring in the ROI.


Figure~\ref{Figure:15} displays the evolution of a few MMFs relevant to the transient brightening events 
described above. To ease comparison, we overplotted the intensity contours of the studied events on all of the maps in Figure~\ref{Figure:15}. 
Close to each footpoint of \textit{E1}, we note type I MMFs (zoomed in the inserts in the left and right corners in Figure~\ref{Figure:15}, panels a, b, c, and d).

The MMF at the eastern side streamed out from the penumbra. Eventually, at  13:24~UT, i.e. at the time the \textit{E1} occurred, 
a unipolar feature remained. The MMF at the western footpoint likely underwent a partial magnetic cancellation of the positive flux (see panel d in Figure~\ref{Figure:15}), slightly before \textit{E1}. After \textit{E1}, we note  two couples of type I MMFs  
(see insert in the left corner of Figure~\ref{Figure:15}, panel e) that disappeared at the time of \textit{E3}. At the western footpoint, the negative flux patch from previous MMF merging was approached by a positive flux feature 
leading to a partial magnetic flux cancellation event.  





The photospheric structures appeared in the penumbra facing the second EFR had vertical magnetic fields. In the same region we found a bright gap and diverging filaments. We consider these features as the counterpart of the positive polarity of the EFR, which broke up the uncombed penumbra. This is likely due to the interaction of rising magnetic flux elements with the inclined magnetic field of the penumbra \citep{Ichimoto2007,Su2010}.

The same regions also exhibited  
LOS velocity opposite to the Evershed flow observed in the penumbral filaments nearby. We consider that these opposite flows are a further consequence of the above change in the field inclination. 
Counter Evershed flows were also found in umbral filaments \citep[e.g.,][]{kleint2013}, during the formation of penumbrae \citep{Schl12,Rom14,Murabito16,Murabito18}, and over stable penumbrae \citep{Louis2014,Siutapia2017}. 


\citet{Lim2011} presented observations of two granular-like features (GLFs), about 1\arcsec{} wide, at the tip of penumbral filaments and associated with MMFs. Both the observed GLFs were preceded by elongated darkening and one of these showed fine structure consisting of thin, dark and bright threads. The two GLFs revealed significant chromospheric transients, specifically brightening in H$\alpha$ and \textit{SDO}/AIA 1600~\AA{} images and jets. The relatively long-lasting, dark patches we observed in the penumbra were 5 times wider than reported by the above authors and showed no threads. Besides, the GLFs found by \citet{Lim2011} had opposite polarity than the field in the sunspot umbra, whereas the dark structure observed in our region had same polarity of the umbral field.

The brightening observed by \citet{Bai2016}, about 10\arcsec{} wide, crossed a penumbra. It started  at the border of the penumbra and then elongated towards the umbra, by leaving signatures detected from the chromosphere to the corona. The transient events analyzed in this work differ from the one reported by \citet{Bai2016} mostly for their larger extension and activation over an area that included both the penumbra and moat, as well as for their longer lifetime. Moreover, they do not show any expansion motion.

\textit{IRIS} spectroscopic information are relevant to the eastern footpoints of the events \textit{E1} and \textit{E3}, embedded in the penumbra. According to \citet{Rathore15}, \textit{IRIS} \ion{C}{2} intensity profiles are a powerful discriminant of upper chromospheric structures, and their asymmetries and line shifts are a robust velocity diagnostic. They showed that the \ion{C}{2} spectral profiles can have a single peak or two or more peaks, depending on how the source function varies with the atmospheric height. They attributed the asymmetry of intensity profiles to velocity gradients between the formation height of the intensity peaks and of the line core. In particular, by comparing the intensity profiles measured in active and quiet Sun regions, \citet{Rathore15} reported line profiles with the blue peak stronger/weaker than the red one and attributed this profile feature to plasma with downflow/upflow above the peak formation height. In our data, we found: i) the intensity profiles measured during the initial stages of both events with red peaks stronger than the blue ones, thus revealing plasma upflow at the time of those measurements; ii) at maximum brightening, intensity profiles with blue peaks stronger than red ones, thus hinting at plasma downflow; iii) \textit{E1} showed a stronger asymmetry of the \ion{C}{2} intensity profiles than derived from \textit{E3};  iv) the intensity profiles measured after \textit{E3}  display blue peaks stronger than red ones, whereas the ones measured after \textit{E1} do not clearly show the same feature. 

The \ion{C}{2} spectra of the brightening event reported by \citet{Bai2016} are asymmetric with the red peak stronger than the blue one, in contrast with the observations analyzed in our study. 

The intensity profiles of the chromospheric \ion{Mg}{2} h\&k lines measured during \textit{E1} and \textit{E3} are also double-peaked but rather symmetric except for the data taken at maximum brightening which have blue peaks stronger than red ones as  for plasma downflow at the formation height of the intensity peaks  \citep{Leenaarts2013}. Finally,  the width of the intensity profiles of the \ion{Mg}{2} h\&k lines does not change significantly during the studied events as for a temperature increase associated to the brightening not occurring in the lower chromosphere \citep{Pereira15}. 
It is worth noting that the \ion{Mg}{2} 2796.35~\AA{} intensity profile presented by \citet{Bai2016} is symmetric as expected from quiet plasma at the formation height of the analyzed intensity peak in contrast to our data.  

Indeed, we notice that the blue peaks are stronger than the red ones  for both the \ion{C}{2} and \ion{Mg}{2} h\&k lines data, especially during event \textit{E3}, which is the stronger brightening. This asymmetry suggests the occurrence of chromospheric evaporation at the penumbral footpoints following a magnetic reconnection process occurring at higher atmospheric heights, as hypothised by \citet{Bai2016}. 

Furthermore, the \ion{Mg}{2} intensity profiles we measured during both \textit{E1} and \textit{E3} show a weak emission of \ion{Mg}{2} 2798.8~\AA{} triplet.  \citet{Pereira15} attributed this rare emission to steep temperature increase in the lower chromosphere. The \ion{Mg}{2} 2798.8~\AA{} emission can involve either the far wings or core of the line, depending on the range of interested column masses. In particular, the emission seen in the wings of intensity profile points to plasma heating occurring deeper down the chromosphere while that recorded at the line core, which is suggestive of pure chromospheric heating \citep{Hansteen17}. Our data show \ion{Mg}{2} 2798.8~\AA{} emission in the line core, thus revealing that \textit{E1} and \textit{E3} produced a steep temperature increase at chromospheric heights.  

We also considered the results presented above with respect to the findings of MHD numerical simulations of coronal and chromospheric micro-flares performed by \citet{Jiang2012}. In these numerical models the micro-flares leave different atmospheric signatures and have different size depending on the atmospheric height of the magnetic reconnection episode responsible for the flaring. In particular, in the case of the simulated coronal micro-flares, the size of the brightening region ranges  from 15\arcsec{} to 22\arcsec. The upper atmosphere response to the event simulated by \citet{Jiang2012} includes both hot plasma jets ($\approx 1.8 \times 10^{6}$~K) observed in the EUV/X-ray bands  and cold jets ($\approx 10^{4}$~K) revealed as H$\alpha$/Ca brightening events. Moreover, the numerical studies indicate a time delay between the signatures of the reconnection process seen in the upper and lower atmosphere. In particular, the enhanced EUV/X-ray emission was found to appear before the H$\alpha$/Ca brightening. \citet{Jiang2012} roughly derived an estimate for this time delay. By taking into account the temperature response at 1500~km (the height formation of H$\alpha$, \citealp{Vernazza81}) those authors reported a time delay of about 3-5 minutes. The spatial scale of the studied brightening events in their signatures in the various layers of atmosphere agree with those reported by the simulation results of \citet{Jiang2012} for coronal micro-flares.

We have seen that the magnetic configuration analyzed in Figure~\ref{Figure:15} for the three penumbral brightening suggest a clear link between the EFR, MMFs and the pre-existing penumbral field. In regard to the trigger mechanism of the observed transient events, \textit{E1} and \textit{E3} seem due to the interaction between the penumbral fields and the EFR. In particular, the outer negative footpoint of the EFR interacts with the MMFs departing from the penumbra, as suggested in the model proposed by \citet{Kano2010} (flux emergence case). Moreover, the positive footpoint of the EFR inside the penumbra also comes into contact with negative polarities around the penumbra originated by either MMFs or flux return patches. This interaction may activate magnetic reconnection events. On the other hand, \textit{E2} seems due to the interplay between MMFs and the pre-existing diffuse field to the south of the sunspot, as suggested by the crossing enhanced connections visible in the \textit{Hinode}/XRT images. This is reminiscent of the model introduced by \citet{Kano2010} (MMFs interaction case). Therefore, for the three studied events there is a tight distinctive interplay among EFR, MMFs and pre-existing magnetic flux systems.

\section{Conclusions} 
In this paper, we analyzed multi-wavelength observations of the AR NOAA 12546 obtained with state-of-the-art instruments to further investigate the trigger mechanism of penumbral brightening events. In particular, we studied the evolution of the AR penumbral area adjacent to two small-scale EFRs emerged in the AR on 2016 May~20. The analyzed region showed three brightening events detected from the chromosphere to the corona. From analysis of IBIS, \textit{SDO}/HMI, \textit{SDO}/AIA, \textit{IRIS}, and \textit{Hinode} observations we found that two of the studied events likely derived from reconnection processes occurred at different atmospheric heights that were activated by the interaction of the pre-existing fields with either the EFRs or MMFs.

Penumbral transient brightening events have been attributed to magnetic reconnection processes occurring in the low corona \citep{Bai2016}. The spatial scale of the events analysed in our study in their signatures in the atmosphere agree with those reported by the simulation results of \citet{Jiang2012} for coronal micro-flares. Moreover, our observations confirm the model for nano-flare events sketched in Figure 12 of the paper by \citet{Kano2010}, both for the flux emergence and MMFs interaction cases. They also complement the results presented by \citet{Bai2016} showing the signatures of the chromospheric evaporation following the flaring events not found by those authors.

We plan to further analyse penumbral brightening events to understand the role of the less-energetic occurrences in the heating of the chromosphere and upper atmosphere. Spectropolarimetric observations to accurately infer the vector magnetic field at different heights in the solar atmosphere, and simultaneous images and spectra 
from the photosphere outward, as the ones expected by the upcoming new generation solar observatories (DKIST, \citealp{Tritschler2015}; Solar Orbiter, \citealp{Muller2013}) are essential to perform these studies.


\acknowledgments
The authors are grateful to Doug Gilliam for the IBIS observations analysed in this study.  
S.L.G and M.M. wish to thanks Peter Young for useful discussion.
This work has received funding from the European Union's Horizon 2020 research and Innovation program under grant agreements No.~739500 (PRE-EST) and No.~824135 (SOLARNET).  
S.L.G.~acknowledges support from the Universit\`a degli Studi di Catania (Piano per la Ricerca Universit\`{a} di Catania 2016-2018 -- Linea di intervento~1 ``Chance''; Linea di intervento~2 ``Ricerca di Ateneo - Piano per la Ricerca 2016/2018'').
This research has made use of NASA's Astrophysics Data System. \textit{IRIS} is a NASA small explorer mission developed and operated by LMSAL with mission operations executed at NASA Ames Research center and major contributions to downlink communications funded by the Norwegian Space Center (NSC, Norway) through an ESA PRODEX contract.

\end{document}